# On the Capacity of Wireless Multicast Networks

Seyed Reza Mirghaderi†, Alireza Bayesteh†, and Amir K. Khandani†

**Abstract**

The problem of maximizing the average rate in a multicast network subject to a coverage constraint (minimum quality of service) is studied. Assuming the channel state information is available only at the receiver side and single antenna nodes, the highest expected rate achievable by a random user in the network, called *expected typical rate*, is derived in two scenarios: hard coverage constraint and soft coverage constraint. In the first case, the coverage is expressed in terms of the outage probability, while in the second case, the expected rate should satisfy certain minimum requirement. It is shown that the optimum solution in both cases (achieving the highest expected typical rate for given coverage requirements) is achieved by an infinite layer superposition code for which the optimum power allocation among the different layers is derived. For the MISO case, a suboptimal coding scheme is proposed, which is shown to be asymptotically optimal, when the number of transmit antennas grows at least logarithmically with the number of users in the network.

## I. Introduction

The widespread application of wireless networks has motivated efficient transmission strategies for different applications. One of these applications is data multicasting where a group of users are interested in receiving the same signal, possibly at different levels of resolution. In a wireless multicast network, a common source is transmitted to $N$ users through a fading channel. In such networks, two criteria are usually studied as measures of performance: *network coverage* (minimum quality of service) and *expected rate* (typical quality of service). In the first criterion, the objective is to provide all the users with a minimum service regardless of their channel qualities. In the second criterion, the average data rate observed by a randomly selected user is considered where users with better channel conditions may receive higher data rates. An example for such networks is scalable video broadcasting in which all the subscribers expect to receive a basic signal, while those with better channel conditions might enjoy a higher resolution.

In [1], the challenges in lossy multicasting are studied from an information theoretical point of view. In this work, for an analog Gaussian source with a bandwidth equal to the channel bandwidth, uncoded transmission is shown to achieve the minimum average end-to-end distortion. The scenario in which the source has a larger bandwidth is studied in [2], where different methods of digital transmission are investigated. In [3], a different approach to broadcasting, called static broadcasting, is proposed. It is assumed that all the users receive the same amount of data from a common source, but with different number of channel uses as determined by their respective channel qualities. The actual transmission time in this scheme depends on the user with the worst channel, and hence, the transmission rate might be very low when the number of users is large.

† Coding & Signal Transmission Laboratory (www.cst.uwaterloo.ca), Dept. of Elec. and Comp. Eng., University of Waterloo, Waterloo, Ontario, Canada, N2L 3G1, Tel: 519-884-8552, Fax: 519-888-4338, e-mail: {smirghad, alireza, khandani} @cst.uwaterloo.ca. Financial supports provided by Nortel, and the corresponding matching funds by the Federal government: Natural Sciences and Engineering Research Council of Canada (NSERC) and Province of Ontario: Ontario Centres of Excellence (OCE) are gratefully acknowledged.

In this work, we consider a wireless multicast network in a quasi-static fading environment with additive Gaussian noise. The objective is to maximize the average performance, while a coverage constraint is satisfied. Average performance is defined as the Quality of Service (QoS) observed by a randomly chosen user (typical user), while the coverage requirement relates to the QoS observed by the user(s) with the worst channel condition(s). We assume that the transmission block is large enough to yield reliable communication. However, averaging over time is not possible because of the delay constraints. In other words, all the symbols within a transmission block experience the same channel gain. The channel state information (CSI) of each user is assumed to be known only at the corresponding receiver. For a channel with the above characteristics, the ergodic capacity is not defined, but the outage capacity is defined as the maximum rate decodable with a given probability [4]. In [5], a broadcast approach for a single user channel with these assumptions is proposed which optimizes the expected decodable rate. We apply “multicast outage capacity” and “expected minimum rate” definitions to characterize coverage in the network. Outage capacity is exploited when we have a hard coverage constraint. In this case, we require that with a given probability, within each transmission block, a specific amount of data is received by all the users. In the soft coverage constraint scenario, we relax the coverage constraint by stating it in terms of the expected minimum rate received by all the users within each block. For both hard and soft coverage constraints, another simultaneous criterion is the maximization of the expected typical rate which is defined as the average rate received by a randomly selected user. In general, there is a tradeoff between these two criteria.

The minimum-service criterion has been studied in [6] for a single user fading channel, assuming CSI is known at the transmitter. In that work, given a service outage constraint for a real-time application, the average rate is maximized for a non real-time application sent on top of it. An adaptive variable rate code is proposed and shown to be optimum in that scenario. Similarly, a minimum rate constrained capacity measure is defined for broadcast channels in [7]. It is shown that the minimum rate region is the ergodic capacity region of a broadcast channel, with an effective noise determined by the minimum rate requirements. Using similar assumptions for the CSI availability, a more general case is considered in [8], where each user specifies its rate constraints in a triplet of maximum rate, minimum rate, and a so-called shortage probability.

In this paper, we use a broadcast model for an unknown quasi-static fading channel [5] and show the optimality of this model in our scenario. The same model is used in [9] to propose a multilevel approach for joint source-channel coding in a SISO channel (assuming the CSI is not available at the transmitter). The broadcast approach in [5] is extended in [10]–[12] for the case that the transmitter has partial channel state information, and in [13] for the case of MIMO block fading channel. [14] combines this broadcast approach with Hybrid Automatic Retransmission Request (HARQ) and shows that this approach results in high throughput and low latency in a point-to-point link. References [15], [16] use the broadcast approach in [5] in a two-hop relay network.

We investigate the performance of the proposed scheme in both SISO and MISO cases. The MISO multicast asymptotic capacity limits are examined in [17], when the CSI is available at the transmitter. It is shown that the adverse effect of having a large number of users can be compensated by increasing the number of transmit antennas. In this work, we study a similar scenario and derive asymptotic capacity results for a large number of transmit antennas.

The rest of this paper is organized as follows: In section II, the system model is introduced. Section III focuses on the virtual broadcast model for an unknown fading channel when the network is delay limited. Sections IV and V discuss multicast networks when we have a single antenna at the transmitter and at each receiver. In section IV, we evaluate the optimum performance of the network in terms of the achievable tradeoff between the expected typical rate and the multicast outage capacity (hard coverage). Section V studies a similar problem of computing the optimum tradeoff, but for a soft coverage constraint where the expected minimum rate is used as the coverage criterion. Section VI investigates the MISO case, where we derive asymptotic capacity results for a large number of transmit antennas. Finally, section VII concludes the paper.

Throughout this paper, we represent the norm of the vectors by $\|.\|$, the conjugate transpose operation by $(.)^\dagger$, and the expectation operation by $E[.]$. The notation "$\log$" is used for the natural logarithm, and rates are expressed in *nats*. We denote $f_y(.)$ and $F_y(.)$ as the probability density function and the cumulative density function of random variable $y$, respectively. Notation $1_{\mathcal{A}}$ is used to define a binary function of $x$ which is equal to $1$ if event $\mathcal{A}$ occurs and $0$ otherwise. For given functions $f(n)$ and $g(n)$, $f(n) = o(g(n))$ is equivalent to $\lim_{n\to\infty} \left|\frac{f(n)}{g(n)}\right| = 0$, and $f(n) = \omega(g(n))$ is equivalent to $\lim_{n\to\infty} \frac{f(n)}{g(n)} = \infty$. We use $A \approx B$ to denote the approximate equality between $A$ and $B$, such that by substituting $A$ by $B$ the validity of the equations is not compromised.

## II. System Model

In this paper, we consider a wireless network broadcasting a common message. In the first part, it is assumed that a single-antenna transmitter transmits a common message to $N$ single-antenna receivers. The received signal at the $i$th receiver, denoted by $y_i$, can be written as

$$y_i = s_i x + n_i, \tag{1}$$

where $x$ is the transmitted signal satisfying an average power constraint of $E[x^2] \leq \mathscr{P}$, $n_i \sim \mathcal{CN}(0,1)$ is the Additive White Gaussian Noise (AWGN), and $s_i \sim \mathcal{CN}(0,1)$ is the channel coefficient from the transmitter to the $i$th receiver. The channel gain $h_i = |s_i|^2$, which is assumed to be constant during a transmission block, has the following Cumulative Distribution Function (CDF):

$$F_i(h) = F(h) = 1 - e^{-h}, \forall i.$$

The *typical channel* of the multicast network is defined as the channel of a randomly selected (typical) user. Since all the channels are i.i.d., the typical channel gain distribution satisfies

$$F_{typ}(h) = F(h) = 1 - e^{-h}. \tag{2}$$

Since all the $N$ channels are Gaussian and they receive a common signal, the coverage requirement is determined by the channel with the lowest gain $h_{mul} = \min_i(h_i)$, which is called the *multicast channel*. Due to the statistical independence of the channels, the gain of the multicast channel has the following distribution

$$\Pr\left\{\min_i(h_i) > h\right\} = \left(\Pr\left\{h_i > h\right\}\right)^N = e^{-Nh}.$$

As a result, we have

$$F_{mul}(h) = 1 - \Pr\left\{\min_i(h_i) > h\right\} = 1 - e^{-Nh}.$$

In this paper, we deal with three quality measures defined as follows:

- *Multicast outage capacity*, $R_\epsilon$, is the rate decodable at the multicast channel with probability $(1-\epsilon)$.
- *Expected multicast rate*, $R_{mul}$, is the average rate decodable at the multicast channel, i.e. $R_{mul} = E[R(h)|h = h_{mul}]$, where $R(h)$ is the decodable rate at the channel state $h$.
- *Expected typical rate*, $R_{ave}$, is the average rate decodable by a randomly selected user, i.e. $R_{ave} = E[R(h)]$.

## III. Broadcast Model for an Unknown Quasi-Static Fading Channel

In [5], it is shown that the expected rate for a receiver with a quasi-static block fading channel, unknown at the transmitter, and a stringent delay constraint, is equivalent to a weighted sum rate of a degraded broadcast channel with infinite number of virtual receivers, each corresponding to a realization of the channel. In this paper, we exploit the same model in a more general fashion. Noting our frequent use of this model, we first study it in more details.

In this work, we assume a block fading channel for all users, where the channel state takes values according to a given probability density function (pdf) (assumed to be exponential) at the start of each block, stays unchanged during the coherence time (block length of the channel), and then changes independently at the start of the subsequent block. The channel state information for each channel is assumed to be available only at the corresponding receiver's side. In this case, if there were no constraint on the decoding delay, coding across different fading blocks would be possible, achieving the so-called ergodic capacity. However, in our model, we impose a decoding delay constraint which restricts the receiver to decode within a period equal to the length of a fading block. Each receiver decodes a fraction of the transmitted data which is supported by its corresponding channel. Hence, for any coding scheme, we have a function $R(h)$ which determines the data rate decoded in channel state $h$. The time average of the rate decoded by a given receiver over infinite number of transmission blocks would be $E_h[R(h)]$ [5].

Consider an infinite number of virtual receivers, such that receiver $RX_h$ is experiencing a fading level between $h$ and $h + \mathrm{d}h$. With these settings, $RX_h$ is receiving all the data received by $RX_{h-\mathrm{d}h}$, in addition to $\mathrm{d}R_h$, where

$$\mathrm{d}R_h = R(h) - R(h - \mathrm{d}h).$$

The virtual receivers introduce a degraded *virtual broadcast network* in which the rate associated with user $RX_h$ is $\mathrm{d}R_h$. From the degraded nature of the Gaussian broadcast channel [18], it follows that $\mathrm{d}R_h \geq 0$. The original receiver corresponds to the virtual receiver $RX_h$ with probability $\eta(h)\mathrm{d}h$, where $\eta(h)$ is the channel gain probability distribution function.

With this interpretation, for a given coding scheme, the distinction between different channels introduced in the previous section, namely *multicast channel* and *typical channel*, is translated to the difference between the probability distribution based on which the original receiver is represented by the virtual receivers. Note that both the *multicast* and *typical* channels correspond to the same *common virtual* *broadcast network*, and the measures defined in the previous section could be written as follows [1]:

$$\begin{aligned} R_{ave} &= \int_0^\infty R(h)\eta(h)\mathrm{d}h \\ &= -(1-F(h))R(h)\big|_0^\infty + \int_0^\infty (1-F(h))\mathrm{d}R_h \\ &\overset{(a)}{=} \int_0^\infty (1-F(h))\mathrm{d}R_h, \end{aligned} \tag{3}$$

$$R_{mul} = \int_0^\infty (1-F_{mul}(h))\mathrm{d}R_h, \tag{4}$$

$$R_\epsilon = R(h_\epsilon) = \int_0^{h_\epsilon} \mathrm{d}R_h, \tag{5}$$

where $h_\epsilon = F_{mul}^{-1}(\epsilon) = -\frac{\log(1-\epsilon)}{N}$. In the above equation, $(a)$ follows from the facts that $R(0) = 0$ and $\lim_{h\to\infty}(1-F(h))R(h) = 0$, where the latter is due to the fact that $1-F(h) = e^{-h}$ and $R(h)$ grows logarithmically with $h$. A similar argument is applied to conclude (4). In the case of the multicast channel in (5), the original receiver has a channel gain less than $h_\epsilon$ with probability $\epsilon$, and hence, the highest decodable rate is $R(h_\epsilon)$, with probability $1-\epsilon$.

As seen above, the performance measures in this work are different weighted sum rates of the *virtual broadcast network*, resulting in vectors $(R_\epsilon, R_{ave})$ and $(R_{mul}, R_{ave})$ for the hard and soft coverage constraint scenarios, respectively. We refer to the maximum achievable regions for these vectors as the *capacity region* corresponding to each case. Next, we provide a definition for the optimality of a performance vector.

**Definition 1** *The boundary set $B$ of a closed convex region $\mathscr{R} \subset \mathbb{R}_+^n$ is defined as*

$$B = \{\mathbf{x} \in \mathscr{R} | \nexists \mathbf{x}' \in \mathbb{R}_{++}^n, \mathbf{x} + \mathbf{x}' \in \mathscr{R}\} \tag{6}$$

*where $\mathbb{R}_+$ and $\mathbb{R}_{++}$ are the set of nonnegative and strictly positive real numbers, respectively.*

With the above definition, a performance vector is optimal if it is in the boundary set of all possible performance vectors. In the following theorem, we show that the optimal performance vector for each coverage constraint scenario is achieved by superposition coding in which the rate of the virtual receiver $RX_h$ is given by

$$\mathrm{d}R_h = \log\left(1 + \frac{h\rho(h)\mathrm{d}h}{1 + h\int_{h+\mathrm{d}h}^\infty \rho(u)\mathrm{d}u}\right), \tag{7}$$

where $\rho(.)$ is the power distribution function such that $\rho(h)\mathrm{d}h$ is the amount of power allocated to the virtual receiver corresponding to the channel state $h$.

[1] With a small misuse of notation, to simplify the problem formulation, the integration in the right hand side of all of the three equations is computed over $h$, where $\mathrm{d}R_h$ is expressed as an explicit function of $h$ throughout the paper.

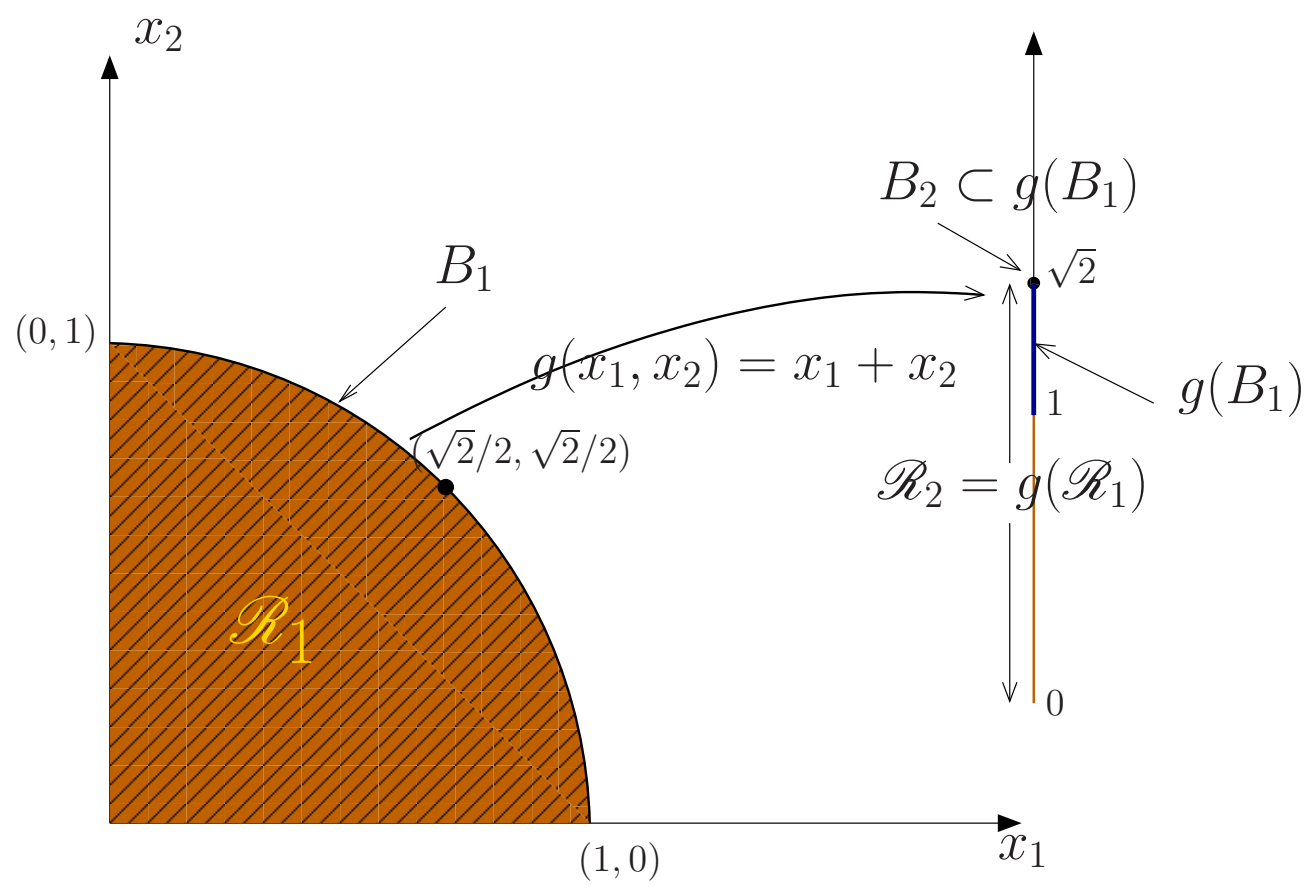

Fig. 1. A schematic figure for Lemma 1.

**Theorem 1** *The boundary set of* $(W_1, ..., W_k)$, *where* $W_i = \int_0^\infty w_i(h)\mathrm{d}R_h \ \forall i$ *is a positive weighted sum rate of the underlying virtual broadcast channel, is achievable by a super-position coding scheme.*

*Proof:* To prove the theorem, we first state and prove the following lemma:

**Lemma 1** *Consider a mapping function* $g(.)$ *from a closed region* $\mathscr{R}_1 \subset \mathbb{R}_+^n$ *to* $\mathscr{R}_2 \subset \mathbb{R}_+^k$ *such that* $g(\mathbf{x}) = \mathbf{M}\mathbf{x}$, *where* $\mathbf{M} \in \mathbb{R}_+^k \times \mathbb{R}_+^n$, *such that each row of* $\mathbf{M}$ *contains at least one positive element. Denote* $B_1$ *and* $B_2$ *as the boundary sets of the regions* $\mathscr{R}_1$ *and* $\mathscr{R}_2$, *respectively. Then, we have*

$$B_2 \subset g(B_1)$$

*Proof:* Assume this is not true. Hence, there must exist $\mathbf{x}_2 \in B_2$ such that $\mathbf{x}_2 \notin g(B_1)$ and $\mathbf{x}_1 \in \mathscr{R}_1$, such that $\mathbf{x}_2 = g(\mathbf{x}_1)$. Since $\mathbf{x}_1 \notin B_1$, there exists $\mathbf{x}_1' \in B_1$ such that $\mathbf{x}_1' - \mathbf{x}_1 \in \mathbb{R}_{++}^n$. Defining $\mathbf{x}_2' = g(\mathbf{x}_1') \in \mathscr{R}_2$, we have

$$\mathbf{x}_2' - \mathbf{x}_2 = \mathbf{M}(\mathbf{x}_1' - \mathbf{x}_1) \subset \mathbb{R}_{++}^k, \tag{8}$$

which contradicts the fact that $\mathbf{x}_2$ is in the boundary set of $\mathscr{R}_2$ and the lemma is proved. ∎

Figure (1) illustrates an example of Lemma 1 when $n = 2$ and $k = 1$. The region $\mathscr{R}_1$ in this example is defined by $x_1, x_2 \geq 0, x_1^2 + x_2^2 \leq 1$ and the mapping is defined by $g(x_1, x_2) = x_1 + x_2$. The boundary of $\mathscr{R}_1$, i.e., $B_1$, is the quarter-circle $x_1, x_2 \geq 0, x_1^2 + x_2^2 = 1$, which is mapped to the line segment $[1, \sqrt{2}]$. The boundary of $\mathscr{R}_2 = g(\mathscr{R}_1)$, i.e., $B_2$, which is the line segment $[0, \sqrt{2}]$, is the point $\sqrt{2}$, which is a subset of $g(B_1)$.

We can arrange the set of the rates of the virtual broadcast network in a rate vector $\mathrm{d}\mathbf{R} = [\mathrm{d}R_h]_0^\infty$ and use the result of Lemma 1, when $n$ tends to infinity. In this case, the matrix transform will tend to $k$ weighted sums of the infinite dimensional vector $\mathbf{x} = \mathrm{d}\mathbf{R}$ as follows:

$$\mathbf{M}\mathbf{x} \rightarrow \left[ \int_0^\infty w_1(h)\mathrm{d}R_h, \cdots, \int_0^\infty w_k(h)\mathrm{d}R_h \right].$$

Setting $x(h) = \mathrm{d}R_h$, we can conclude that the boundary region of $(W_1, \cdots, W_k)$ is a subset of the transformation of the boundary set of the rate vector $\mathrm{d}\mathbf{R}$, which is achieved, as shown in [19], by superposition coding. In other words, any vector $\mathbf{v}$ in the boundary set of $(W_1, \cdots, W_k)$, is achieved by a multi-layer code in which the rate of the virtual receiver $RX_h$ is equal to

$$\log\left(1 + \frac{h\rho_{\mathbf{v}}(h)\mathrm{d}h}{1 + h\int_{h+\mathrm{d}h}^{\infty}\rho_{\mathbf{v}}(u)\mathrm{d}u}\right),$$

where the power distribution function $\rho_{\mathbf{v}}(h)$, indexed with $\mathbf{v}$ to emphasize its dependence on this vector, satisfies $\int_0^{\infty}\rho_{\mathbf{v}}(u)\mathrm{d}u = \mathscr{P}$. This completes the proof of Theorem 1. ■

Using the above theorem, it easily follows that the optimal performance vectors $(R_\epsilon, R_{ave})$ and $(R_{mul}, R_{ave})$, defined in the sections IV and V, respectively, can be achieved using superposition coding.

## IV. Hard Coverage Constraint

In this section, we consider a scenario where a minimum rate of $R_\epsilon$ (multicast outage capacity) should be delivered to all the users with a probability of $(1 - \epsilon)$, where $\epsilon$ is the probability of outage. Given this constraint, we want to maximize the average rate received by a randomly chosen user in the network, i.e. $R_{ave}$. This includes the rate received by a typical user, even if the user is in outage. We can categorize the possible states of the *virtual broadcast network* in two group: (i) $h \leq h_\epsilon$, in which case the rate $\mathrm{d}R_h$ contributes to both $R_\epsilon$ and $R_{ave}$, and (ii) $h > h_\epsilon$, in which case the rate $\mathrm{d}R_h$ contributes only to $R_{ave}$.

The problem of maximizing $R_{ave}$ without any constraint on $R_\epsilon$ is studied in [5]. On the other hand, from [4], we know that the maximum $R_\epsilon$ without any constraint on $R_{ave}$ is achieved by a single layer code with power $\mathscr{P}$ and rate

$$C_\epsilon = \log(1 + h_\epsilon\mathscr{P}). \tag{9}$$

In this section, we investigate the tradeoff between $R_\epsilon$ and $R_{ave}$.

Setting $w_1(h) = 1_{\{h \leq h_\epsilon\}}$ and $w_2(h) = 1 - F(h)$, Theorem 1 states that the boundary set of $(R_\epsilon, R_{ave})$ is achieved by superposition coding, in which

$$\mathrm{d}R_h = \log\left(1 + \tfrac{h\rho(h)\mathrm{d}h}{1+hI(h)}\right) = \int_{I(h)}^{I(h)+\rho(h)\mathrm{d}h} \tfrac{h\mathrm{d}p}{1+hp}, \tag{10}$$

and $I(h) = \int_{h+\mathrm{d}h}^{\infty}\rho(u)\mathrm{d}u$. Note that $\mathrm{d}R_h$ is not necessarily very small since our power allocation function might have some impulses. As stated earlier, we want to jointly optimize the weighted sum of these rates according to the weighting functions $w_1(h)$ and $w_2(h)$. The optimization is on the function $\rho(h)$, however, in the following we show that it can be simplified to a point optimization problem.

**Definition 2** *The channel gain-interference function, $s(p)$ is defined as*

$$s(p) \triangleq \sup \{h| \ I(h) \geq p\} . \tag{11}$$

In fact, the *channel gain-interference function* $s(p)$ is the inverse of the interference function in terms of the channel gain in the points it is invertible and determines the channel gain of the virtual receiver experiencing the interference level $p$. It is evident that $s(p)$ is a decreasing function of $p$. According to (10), we can write the expected typical rate as

$$R_{ave} = \int_0^{\infty}(1 - F(h))\mathrm{d}R_h = \int_0^{\mathscr{P}} g(p, s(p))\mathrm{d}p, \tag{12}$$

where $g(x, y) = (1 - F(y))\frac{y}{1+xy}$. In deriving (12), we have used the fact that the contribution of the virtual receiver $RX_h$ in the expected typical rate, from (3), can be written as $(1 - F(h))\mathrm{d}R_h = (1 - F(h)) \int_{I(h)}^{I(h)+\rho(h)\mathrm{d}h} \frac{h\mathrm{d}p}{1+hp}$. Noting that in the interval $p \in [I(h), I(h)+\rho(h)\mathrm{d}h]$, $h$ can be written as $s(p)$ [2], the contribution of $RX_h$ in $R_{ave}$ can be written as $\int_{I(h)}^{I(h)+\rho(h)\mathrm{d}h} g(p, s(p))\mathrm{d}p$. Since $R_{ave}$ is the summation of the contributions of all virtual receivers, we have $R_{ave} = \int_h \int_{I(h)}^{I(h)+\rho(h)\mathrm{d}h} g(p, s(p))\mathrm{d}p = \int_{I(0)}^{I(\infty)} g(p, s(p))\mathrm{d}p = \int_0^{\mathscr{P}} g(p, s(p))\mathrm{d}p$, which is due to the fact that $I(0) = \mathscr{P}$ and $I(\infty) = 0$.

Differentiating $g(x, y)$ with respect to $y$, we obtain

$$\frac{\partial}{\partial y} g(x, y) = \frac{1 - F(y) - yf(y)(1 + xy)}{(1 + xy)^2}. \tag{13}$$

In the case of Rayleigh fading, we have $f(y) = 1 - F(y) = e^{-y}$. By studying the behavior of $\frac{\partial}{\partial y} g(x, y)$, it is easy to show that $\frac{\partial}{\partial y} g(x, y)|_{x=p}$ is positive for $y < I_0^{-1}(p)$ and is negative for $y > I_0^{-1}(p)$, where $I_0(h) = \frac{(1-F(h))-hf(h)}{h^2 f(h)}$. As a result,

$$\arg\max(g(x, y)|_{x=p}) = I_0^{-1}(p). \tag{14}$$

Note that $I_0(.)$ [3] is indeed the interference function corresponding to the optimal power allocation in the unconstrained maximization of $R_{ave}$ solved in [5].

**Definition 3** *The multicast interference level $\alpha$ for a given channel gain-interference function $s(.)$ is defined as*

$$\alpha \triangleq \min \{p|s(p) \leq h_\epsilon\} . \tag{15}$$

In fact, $\alpha$ is the level of the interference observed by all the virtual receivers contributing to $R_\epsilon$ due to the power allocated to the upper levels not contributing to $R_\epsilon$. It follows from definitions 2 and 3 that

[2] Note that this is true even if $\rho(.)$ contains an impulse at $h$.

[3] Note that the function $I_0(.)$ is monotonically decreasing and as a result, it is invertible.

$\alpha = I(h_\epsilon)$. Using this definition and the same arguments as in (12), $R_\epsilon$ in (5) can be written as

$$\begin{aligned} R_\epsilon &= \int_0^{h_\epsilon} \mathrm{d}R_h \\ &= \int_0^{h_\epsilon} \int_{I(h)}^{I(h)+\rho(h)\mathrm{d}h} \frac{h\mathrm{d}p}{1+hp} \\ &= \int_0^{h_\epsilon} \int_{I(h)}^{I(h)+\rho(h)\mathrm{d}h} \frac{s(p)\mathrm{d}p}{1+s(p)p} \\ &= \int_{I(0)}^{I(h_\epsilon)} \frac{s(p)\mathrm{d}p}{1+s(p)p} \\ &= \int_\alpha^{\mathscr{P}} m(p, s(p))\mathrm{d}p, \qquad (16) \\ &= \int_0^{\mathscr{P}} 1_{\{s(p)\le h_\epsilon\}} m(p, s(p))\mathrm{d}p \qquad (17) \end{aligned}$$

where

$$m(x, y) = \frac{y}{1+xy}, \tag{18}$$

and (17) follows from the definition of $\alpha$ in (15). Here, we assume that $h_\epsilon \le 1$. This is not a restricting assumption as the solution in the other case ($h_\epsilon > 1$) follows using a similar approach. However, this assumption simplifies the derivations as it guarantees $I_0(h_\epsilon) > 0$. On the other hand, this assumption is equivalent to $\epsilon \le 1 - e^{-N}$, which covers most of the cases of interest since we expect the outage probability to be much smaller than 1. Using (12) and (16), the problem is translated to

$$\max_{s(.)} R_{ave} = \max_{s(.)} \int_0^{\mathscr{P}} g(p, s(p))\mathrm{d}p, \tag{19}$$

subject to

$$R_\epsilon = \int_\alpha^{\mathscr{P}} m(p, s(p))\mathrm{d}p \ge \zeta C_\epsilon, \tag{20}$$

where $C_\epsilon$ is defined in (9), and $\zeta$ is a normalization factor which expresses $R_\epsilon$ in terms of $C_\epsilon$. Since it is not possible to achieve values of $R_\epsilon$ above $C_\epsilon$, we can restrict ourselves to the values of $\zeta$ between $0$ and $1$. Note that the maximization in (19) is over all decreasing positive functions $s(.)$. Also note that $\alpha$ in (20) is the *multicast interference level* defined in (15) and depends on the function $s(.)$.

From (18), we note that for any chosen $x$, $m(x, y)$ is an increasing function of $y$. Hence, noting the definition of $\alpha$ in (15), we can write

$$\begin{aligned} R_\epsilon \le \int_\alpha^{\mathscr{P}} m(p, h_\epsilon)\mathrm{d}p &= \log\left(\frac{1+h_\epsilon \mathscr{P}}{1+h_\epsilon \alpha}\right) \\ &= C_\epsilon - \log(1+h_\epsilon \alpha). \qquad (21) \end{aligned}$$

Therefore, following (9), (20) and (21), we have

$$\begin{aligned} \zeta C_\epsilon &\leq C_\epsilon - \log(1 + h_\epsilon \alpha) \\ \Rightarrow \alpha &\leq \frac{e^{(1-\zeta)C_\epsilon} - 1}{h_\epsilon}. \end{aligned} \tag{22}$$

**Lemma 2** *For the optimizer of (19), we have* $\alpha \leq I_0(h_\epsilon)$.

*Proof:* Assume $\alpha > I_0(h_\epsilon)$. Denote the optimizer function by $s^*(.)$ and its resulting multicast outage capacity and expected typical rate by $R^*_\epsilon$ and $R^*_{ave}$, respectively. Also, define $\hat{s}(p)$ as

$$\hat{s}(p) = \begin{cases} I_0^{-1}(p) & p < I_0(h_\epsilon) \\ h_\epsilon & I_0(h_\epsilon) \leq p \leq \alpha \\ s^*(p) & p > \alpha \end{cases}, \tag{23}$$

and its resulting multicast outage capacity and expected typical rate by $\hat{R}_\epsilon$ and $\hat{R}_{ave}$, respectively. We can write

$$\begin{aligned} &\hat{R}_\epsilon - R^*_\epsilon \overset{(17)}{=} \\ &\int_0^{\mathscr{P}} \left[1_{\{\hat{s}(p) \leq h_\epsilon\}} m(p, \hat{s}(p)) - 1_{\{s^*(p) \leq h_\epsilon\}} m(p, s^*(p))\right] \mathrm{d}p \overset{(a)}{=} \\ &\int_{I_0(h_\epsilon)}^{\alpha} m(p, \hat{s}(p)) \mathrm{d}p > 0, \end{aligned} \tag{24}$$

where $(a)$ follows from the facts that (i) $\hat{s}(p) \leq h_\epsilon$ for $p \geq I_0(h_\epsilon)$, (ii) $s^*(p) \leq h_\epsilon$ for $p \geq \alpha$ (from the definition of $\alpha$ in (15)), and (iii) for $p \geq \alpha$, $\hat{s}(p) = s^*(p)$ (from (23)). (24) implies that $\hat{R}_\epsilon > R^*_\epsilon$. Moreover, we have

$$\begin{aligned} \hat{R}_{ave} - R^*_{ave} &= \int_0^{\mathscr{P}} (g(p, \hat{s}(p)) - g(p, s^*(p))) \mathrm{d}p \\ &\overset{(a)}{=} \int_0^{I_0(h_\epsilon)} \left[g(p, I_0^{-1}(p)) - g(p, s^*(p))\right] \mathrm{d}p \\ &+ \int_{I_0(h_\epsilon)}^{\alpha} (g(p, h_\epsilon) - g(p, s^*(p))) \mathrm{d}p > 0, \end{aligned} \tag{25}$$

where $(a)$ follows from the fact that for $p \geq \alpha$, $\hat{s}(p) = s^*(p)$. In the above inequality, the positivity of the first term in the left hand side is concluded from (14), and the positivity of the second term is concluded from the fact that $s^*(p) > h_\epsilon$ for $p \leq \alpha$ (from the definition of $\alpha$ in (15)), and also (13) which implies that $g(x, y)|_{x=p}$ is decreasing for $y > I_0^{-1}(p)$. Therefore, $\hat{R}_{ave} > R^*_{ave}$ and $\hat{R}_\epsilon > R^*_\epsilon$, which contradict our assumption of optimality of $s^*(.)$ and the lemma is proved. ■

The above lemma states the fact that, applying the multicast outage constraint, more power will be allocated to the channel gains lower than the outage threshold, compared to the unconstrained scenario studied in [5].

**Lemma 3** *Given $\alpha$, the optimizer of (19) is given by*

$$s_\alpha(p) = \begin{cases} I_0^{-1}(p) & p < \alpha \\ h_\epsilon & \alpha \le p \le I_\lambda^{hc}(h_\epsilon) \\ I_\lambda^{hc^{-1}}(p) & p > I_\lambda^{hc}(h_\epsilon) \end{cases}, \tag{26}$$

*where $I_\lambda^{hc}(h) = \frac{(\lambda+1-F(h))-hf(h)}{h^2 f(h)}$ in which the superscript $(.)^{hc}$ stands for the "hard coverage" constraint scenario, $I_\lambda^{hc^{-1}}(.)$ represents the inverse of the function $I_\lambda^{hc}(.)$, and*

$$\lambda = \begin{cases} 0, & if \int_\alpha^{\mathscr{P}} m(p, s_\alpha(p))\mathrm{d}p > \zeta C_\epsilon \\ \arg\left(\int_\alpha^{\mathscr{P}} m(p, s_\alpha(p))\mathrm{d}p = \zeta C_\epsilon\right), & otherwise \end{cases}.$$

*Proof:* As observed from (20), the value of $s(p)$ in the range $0 \le p \le \alpha$ dose not affect the multicast constraint. Hence, (19) can be written as

$$\begin{aligned} \max_{s(.)} R_{ave} &= \max_{s(.)} \int_0^\alpha g(p, s(p))\mathrm{d}p \\ &+ \max_{s(.), \int_\alpha^{\mathscr{P}} m(p,s(p))\mathrm{d}p \ge \zeta C_\epsilon} \int_\alpha^{\mathscr{P}} g(p, s(p))\mathrm{d}p \end{aligned} \tag{27}$$

Note that as mentioned earlier, all the maximizations are performed over positive decreasing functions. Moreover, since $s(\alpha) = h_\epsilon$ (due to the definition of $\alpha$ in (15)), the solution of the above maximization problem must satisfy $s(p) \ge h_\epsilon$ in the interval $p \in [0, \alpha]$ and $s(p) \le h_\epsilon$ elsewhere. (14) implies that the first term in (27) can be upper-bounded as follows:

$$\max_{s(.)} \int_0^\alpha g(p, s(p))\mathrm{d}p \le \int_0^\alpha g(p, I_0^{-1}(p))\mathrm{d}p. \tag{28}$$

Moreover, writing K.K.T. condition for the second term of (27), the maximization problem will be translated to

$$\max_{s(.)} \int_\alpha^{\mathscr{P}} T_\lambda(p, s(p))\mathrm{d}p, \tag{29}$$

where $T_\lambda(x, y) = g(x, y) + \lambda m(x, y)$. $\lambda$ is $0$, if the outage constraint is not limiting; otherwise, its value can be computed using the outage constraint $\int_\alpha^{\mathscr{P}} m(p, s(p))\mathrm{d}p = \zeta C_\epsilon$. Differentiating the function $T_\lambda(x, y)$ with respect to $y$, we obtain

$$\frac{\partial}{\partial y} T_\lambda(x, y) = \frac{\lambda + 1 - F(y) - yf(y)(1 + xy)}{(1 + xy)^2}. \tag{30}$$

In the case of Rayleigh fading, by studying the behavior of $\frac{\partial}{\partial y} T_\lambda(x, y|_{x=p})$, it follows that $\frac{\partial}{\partial y} T_\lambda(x, y|_{x=p}) > 0$ for $y < I_\lambda^{hc^{-1}}(p)$ and $\frac{\partial}{\partial y} T_\lambda(x, y|_{x=p}) < 0$ for $y > I_\lambda^{hc^{-1}}(p)$, where $I_\lambda^{hc}(h) = \frac{\lambda+1-F(h-hf(h)}{h^2 f(h)}$ [4]. As a result,

$$\arg\max T_\lambda(x, y)|_{x=p} = I_\lambda^{hc^{-1}}(p). \tag{31}$$

[4]Note that the function $I_\lambda^{hc}(.)$ is monotonically decreasing for $h < h_\epsilon$ (which is the region of interest here), and as a result, it is invertible.

Note that as $I_0(h) \leq I_\lambda(h)$, $\forall h$, and $\alpha \leq I_0(h_\epsilon)$ (from Lemma 2), it follows that $\alpha \leq I_\lambda^{hc}(h_\epsilon)$. Hence, the integral $\int_\alpha^{\mathscr{P}} T_\lambda(p, s(p))\mathrm{d}p$ can be written as the summation of the two integrals $\int_\alpha^{I_\lambda^{hc}(h_\epsilon)} T_\lambda(p, s(p))\mathrm{d}p$ and $\int_{I_\lambda^{hc}(h_\epsilon)}^{\mathscr{P}} T_\lambda(p, s(p))\mathrm{d}p$. First, we note that in the whole interval of $[\alpha, \mathscr{P}]$, we have $s(p) \leq h_\epsilon$ (due to the definition of $\alpha$). Moreover, as $I_\lambda^{hc^{-1}}(.)$ is a decreasing function (in the interval $[\alpha, \mathscr{P}]$), we have $h_\epsilon \leq I_\lambda^{hc^{-1}}(p)$, in the interval $[\alpha, I_\lambda^{hc}(h_\epsilon)]$. Combining this with the fact that $s(p) \leq h_\epsilon$ in $[\alpha, I_\lambda^{hc}(h_\epsilon)]$ implies that $s(p) \leq I_\lambda^{hc^{-1}}(p)$ in this interval. As seen before, $T_\lambda(x, y)$ is an increasing function of $y$ for $y < I_\lambda^{hc^{-1}}(x)$. Consequently,

$$\int_\alpha^{I_\lambda^{hc}(h_\epsilon)} T_\lambda(p, s(p))\mathrm{d}p \leq \int_\alpha^{I_\lambda^{hc}(h_\epsilon)} T_\lambda(p, h_\epsilon)\mathrm{d}p. \tag{32}$$

Moreover, using (31), we have

$$\int_{I_\lambda^{hc}(h_\epsilon)}^{\mathscr{P}} T_\lambda(p, s(p))\mathrm{d}p \leq \int_{I_\lambda^{hc}(h_\epsilon)}^{\mathscr{P}} T_\lambda\left(p, I_\lambda^{hc^{-1}}(p)\right)\mathrm{d}p. \tag{33}$$

Hence, for any function $s(p)$ such that $s(p) \leq h_\epsilon$ for $p > \alpha$, we can write

$$\int_\alpha^{\mathscr{P}} T_\lambda(p, s(p))\mathrm{d}p \leq \int_\alpha^{\mathscr{P}} T_\lambda(p, s_\alpha(p))\mathrm{d}p. \tag{34}$$

Combining the above equation with (28) yields that

$$\max_{s(.)} R_{ave} \leq \int_0^{\mathscr{P}} g(p, s_\alpha(p))\mathrm{d}p. \tag{35}$$

To complete the proof, it is sufficient to show that $s_\alpha(.)$ satisfies the conditions mentioned earlier; i.e., $s_\alpha(.)$ is a decreasing function and $s_\alpha(\alpha) = h_\epsilon$. The latter is obvious from the definition of $s_\alpha(.)$ in (26). For showing the former, we first note that $I_0^{-1}(.)$ and $I_\lambda^{hc^{-1}}(.)$ are decreasing functions in the intervals $[0, \alpha]$ and $[I_\lambda^{hc}(h_\epsilon), \mathscr{P}]$, respectively. Moreover, from Lemma 2, we have $\alpha \leq I_0(h_\epsilon)$ which implies that $I_0^{-1}(\alpha) \geq h_\epsilon$. This shows that $s_\alpha(.)$ is decreasing in the whole interval $[0, \mathscr{P}]$, which completes the proof of the lemma. ■

An interesting consequence of Lemma 3 is that the problem of maximization over the function $s(.)$ is simplified to the point optimization problem over the value of $\alpha$.

**Theorem 2** *The capacity region of a Rayleigh fading multicast network with a hard coverage constraint is given by*

$$R_{ave} \leq \max_{0 \leq \alpha \leq \min\left(\frac{e^{(1-\zeta)C_\epsilon}-1}{h_\epsilon}, I_0(h_\epsilon)\right)} \int_0^{\mathscr{P}} g(p, s_\alpha(p))\mathrm{d}p, \tag{36}$$

$$R_\epsilon \leq \zeta C_\epsilon, \tag{37}$$

*where $\zeta$ changes from $0$ to $1$.*

*Proof:* The proof follows from Lemma 2, Lemma 3, and inequality (22). ■

**Corollary 1** *For any outage probability* $\epsilon > 0$ *such that* $h_\epsilon \leq I_0^{-1}(\mathscr{P})$*, the capacity region of a Rayleigh fading multicast network, i.e.,* $(R_\epsilon, R_{ave})$*, is given by*

$$R_\epsilon \leq \log\left(1 + \frac{h_\epsilon \beta \mathscr{P}}{1 + h_\epsilon (1-\beta)\mathscr{P}}\right), \tag{38}$$

$$\begin{aligned} R_{ave} &\leq 2(E_1(\theta(\beta)) - E_1(1)) - (e^{-\theta(\beta)} - e^{-1}) \\ &+ e^{-h_\epsilon} \log\left(1 + \frac{h_\epsilon \beta \mathscr{P}}{1 + h_\epsilon (1-\beta)\mathscr{P}}\right), \end{aligned} \tag{39}$$

*where* $\beta$ *changes from* $0$ *to* $1$, $\theta(\beta) = \frac{2}{1+\sqrt{1+4(1-\beta)\mathscr{P}}}$*, and* $E_1(x) \triangleq \int_x^\infty \frac{e^{-t}}{t} \mathrm{d}t$.

*Proof:* Since $h_\epsilon \leq I_0^{-1}(\mathscr{P})$, it follows that $I_0(h_\epsilon) \geq \mathscr{P}$. Noting that $I_0(h) \leq I_\lambda^{hc}(h)$, $\forall h, \lambda \geq 0$, it follows that $I_\lambda^{hc}(h_\epsilon) > \mathscr{P}$ for any $\lambda \geq 0$. Therefore, (26) can be written as

$$s_\alpha(p) = \begin{cases} I_0^{-1}(p) & p < \alpha \\ h_\epsilon & \alpha \leq p \leq \mathscr{P} \end{cases}. \tag{40}$$

In this case, the solution to the maximization problem (36) is $\alpha = \frac{e^{(1-\zeta)C_\epsilon} - 1}{h_\epsilon}$. Defining $\beta \triangleq 1 - \frac{e^{(1-\zeta)C_\epsilon} - 1}{h_\epsilon \mathscr{P}}$, first we note that $0 \leq \beta \leq 1$. Moreover, (37) can be written as

$$\begin{aligned} R_\epsilon &\leq \zeta C_\epsilon \\ &= C_\epsilon - \log\left(1 + h_\epsilon (1-\beta)\mathscr{P}\right) \\ &= \log\left(1 + \frac{h_\epsilon \beta \mathscr{P}}{1 + h_\epsilon (1-\beta)\mathscr{P}}\right). \end{aligned} \tag{41}$$

The interference function corresponding to $s_\alpha(p)$ in (40) can be expressed as

$$\begin{aligned} I(h) &= \alpha U\left(I_0^{-1}(\alpha) - h\right) + \\ &\quad \beta \mathscr{P} U(h_\epsilon - h) + I_0(h) U\left(h - I_0^{-1}(\alpha)\right), \end{aligned} \tag{42}$$

where $U(.)$ denotes the unit step function and $\alpha = (1-\beta)\mathscr{P}$. Differentiating $I(h)$ with respect to $h$ results in the following power allocation function:

$$\rho(h) = (\mathscr{P} - \alpha)\delta(h - h_\epsilon) + \rho_0(h),$$

where

$$\rho_0(h) = \begin{cases} \frac{2}{h^3} - \frac{1}{h^2} & I_0^{-1}(\alpha) < h < 1 \\ 0 & else \end{cases}$$

is the power allocation function in the unconstraint problem [5]. Using (3) and (10), the expected typical rate can be written as

$$\begin{aligned} R_{ave} &= \int_0^\infty (1 - F(h))\mathrm{d}R_h \\ &= e^{-h_\epsilon} \log\left(1 + \frac{h_\epsilon \beta \mathscr{P}}{1 + h_\epsilon (1-\beta)\mathscr{P}}\right) + \\ &\quad \int_{I_0^{-1}(\alpha)}^1 e^{-h} \frac{h\rho_0(h)}{1 + hI_0(h)} \mathrm{d}h \\ &= e^{-h_\epsilon} \log\left(1 + \frac{h_\epsilon \beta \mathscr{P}}{1 + h_\epsilon (1-\beta)\mathscr{P}}\right) + \\ &\quad \int_{I_0^{-1}(\alpha)}^1 e^{-h} \frac{h\left(\frac{2}{h^3} - \frac{1}{h^2}\right)}{1 + h(\frac{1}{h^2} - \frac{1}{h})} \mathrm{d}h \\ &= e^{-h_\epsilon} \log\left(1 + \frac{h_\epsilon \beta \mathscr{P}}{1 + h_\epsilon (1-\beta)\mathscr{P}}\right) + \\ &\quad \int_{I_0^{-1}(\alpha)}^1 2\frac{e^{-h}}{h} \mathrm{d}h - \int_{I_0^{-1}(\alpha)}^1 e^{-h} \mathrm{d}h. \end{aligned} \tag{43}$$

Noting that $I_0^{-1}(\alpha) = \theta(\alpha) = \frac{2}{1+\sqrt{1+4\alpha}}$ and $\alpha = (1-\beta)\mathscr{P}$ completes the proof of the corollary. ■

An interesting conclusion of Corollary 1 is that, the expected typical rate is maximized when the multicast minimum rate is provided in a single layer. In the case we have no multicast constraint, it is shown in [5] that a multi-layer code with a small rate in each layer is optimal in terms of maximizing the expected rate. However, when we are constrained to distribute a fraction of the available power to a set of *low* channel gains $[0, h_\epsilon]$ (coverage constraint), the optimum solution allocates all the power to the highest gain, i.e. $h_\epsilon$.

Note that the assumption $h_\epsilon \le I_0^{-1}(\mathscr{P})$ is not difficult to satisfy, since the outage probability $\epsilon$ is usually small. Moreover, as $h_\epsilon = -\frac{\log(1-\epsilon)}{N}$, the value of $h_\epsilon$ decreases significantly with the number of users. For example, for $N = 5$ and $\mathscr{P} = 100$, the outage probability $\epsilon$ could be as high as $0.38$ in order to have $h_\epsilon \le I_0^{-1}(\mathscr{P})$. Figure (2) shows the capacity region of this network when $\epsilon = 0.01$. It is evident that due to the hard coverage constraint, the multicast outage capacity is in general very small in comparison with the expected typical rate.

## V. Soft Coverage Constraint

In the previous section, we observed that a strict coverage constraint can result in very small values for the multicast outage capacity. We can relax the coverage requirement by relying on the *expected multicast rate*, i.e. $R_{mul}$. This results in the performance vector $(R_{mul}, R_{ave})$ and its corresponding capacity region. According to Theorem 1, the optimality of superposition coding is concluded for this performance vector.

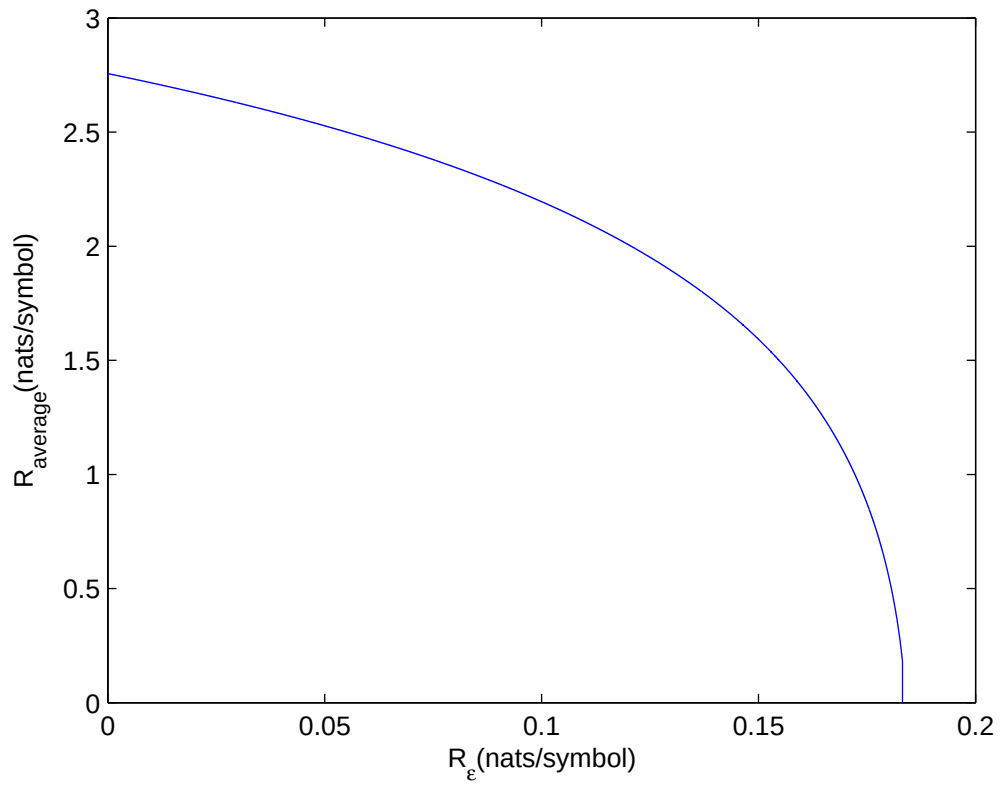

Fig. 2. Hard coverage constraint: multicast outage capacity vs. expected typical rate for $\mathscr{P} = 100$ and $N = 5$.

**Theorem 3** *The capacity region of a Rayleigh fading multicast network with soft coverage constraint is given by*

$$R_{ave} = \int_0^\infty e^{-u} \frac{u\rho_\gamma^{sc}(u)\mathrm{d}u}{1 + uI_\gamma^{sc}(u)} \quad (44)$$

$$R_{mul} = \int_0^\infty e^{-Nu} \frac{u\rho_\gamma^{sc}(u)\mathrm{d}u}{1 + uI_\gamma^{sc}(u)}, \quad (45)$$

*where*

$$I_\gamma^{sc}(h) = \begin{cases} \mathscr{P} & \text{if } h < h_0 \\ \frac{e^{-h}(1-h)+\gamma e^{-Nh}(1-Nh)}{h^2(e^{-h}+\gamma N e^{-Nh})} & h_0 < h < h_1 \\ 0 & h > h_1 \end{cases}, \quad (46)$$

*in which the superscript $(.)^{sc}$ stands for the "soft coverage" constraint scenario, $\rho_\gamma^{sc}(h) = -\frac{\partial I_\gamma^{sc}(h)}{\partial h}$, and $h_0$ and $h_1$ are real numbers, such that*

$$\frac{e^{-h_0}(1-h_0)+\gamma e^{-Nh_0}(1-Nh_0)}{h_0^2(e^{-h_0}+\gamma N e^{-Nh_0})} = \mathscr{P},$$

$$\frac{e^{-h_1}(1-h_1)+\gamma e^{-Nh_1}(1-Nh_1)}{h_1^2(e^{-h_1}+\gamma N e^{-Nh_1})} = 0,$$

*for different values of $\gamma \geq 0$.*

*Proof:* If we set $w_1(h) = 1 - F_{typ}(h)$ and $w_2(h) = 1 - F_{mul}(h)$, Theorem 1 states that we should search for an infinite layer superposition code. Considering $\rho(h)\mathrm{d}h$ as the power of the layer associated with the channel gain $h$, the corresponding rate is[5]

$$\mathrm{d}R_h = \log\left(1 + \frac{h\rho(h)\mathrm{d}h}{1 + hI(h)}\right) = \frac{h\rho(h)\mathrm{d}h}{1 + hI(h)}, \quad (47)$$

[5]Note that here, unlike the hard coverage constraint scenario, the power distribution function does not have any impulses. In fact, the optimization problem in the soft coverage constraint scenario, as seen in the proof, is similar to that of the unconstrained scenario [5] for which the optimal power allocation function has been proved to have no impulses.

where

$$I(h) = \int_h^\infty \mathrm{d}u \rho(u),$$

and

$$I(0) = \mathscr{P}.$$

Using the above expression, the rate corresponding to the fading level $h$ is

$$R(h) = \int_0^h \frac{u\rho(u)\mathrm{d}u}{1 + uI(u)}.$$

Following the definitions, we have

$$R_{mul} = \int_0^\infty (1 - F_{mul}(u))\mathrm{d}R(u) = \int_0^\infty e^{-Nu} \frac{u\rho(u)\mathrm{d}u}{1 + uI(u)}, \tag{48}$$

$$R_{ave} = \int_0^\infty (1 - F_{typ}(u))\mathrm{d}R(u) = \int_0^\infty e^{-u} \frac{u\rho(u)\mathrm{d}u}{1 + uI(u)}. \tag{49}$$

The problem is that given $R_{mul} = r$, what is the maximum achievable $R_{ave}$. In other words,

$$R_{ave} = \max_{I(u)} \int_0^\infty e^{-u} \frac{u\rho(u)\mathrm{d}u}{1 + uI(u)}, \tag{50}$$

subject to:

$$\int_0^\infty e^{-Nu} \frac{u\rho(u)\mathrm{d}u}{1 + uI(u)} = r, \tag{51}$$

$$I(0) = \mathscr{P},$$

and

$$I(\infty) = 0.$$

Equivalently, to derive the capacity region $(R_{mul}, R_{ave})$, it is sufficient to solve the following optimization problem: [6]

$$\max_{I(.)} R_{ave} + \gamma R_{mul}, \tag{52}$$

subject to

$$I(0) = \mathscr{P},$$

$$I(\infty) = 0,$$

[6]Note that as we are allowed to user time-sharing, the capacity region is convex. As a result, the capacity region can be characterized as (52).

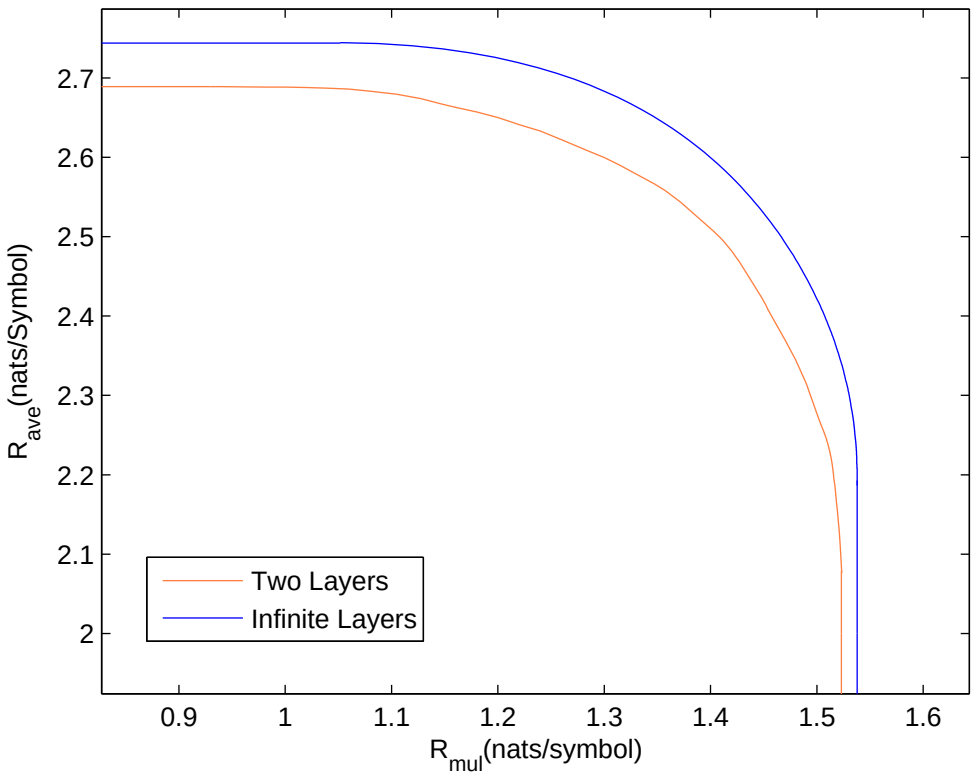

Fig. 3. Soft coverage constraint: expected minimum rate vs. expected typical rate for $\mathscr{P} = 100$ and $N = 5$

for all values of $\gamma \geq 0$. To solve the above optimization problem, we define $S(x, I(x), I'(x), \gamma)$ as follows:

$$S(x, I(x), I'(x), \gamma) = e^{-x} \frac{xI'(x)}{1 + xI(x)} + \gamma e^{-Nx} \frac{xI'(x)}{1 + xI(x)}. \tag{53}$$

Note that

$$I'(x) = -\rho(x).$$

The necessary condition for $I(x)$ to maximize (50) with the constraint (51) is the zero functional variation [20] of $S(x, I(x), I'(x), \gamma)$,

$$\frac{\partial}{\partial I} S - \frac{\mathrm{d}}{\mathrm{d}x} \frac{\partial}{\partial I'} S = 0, \tag{54}$$

where

$$\frac{\partial}{\partial I} S = (e^{-x} + \gamma e^{-Nx}) \frac{x^2 I'(x)}{(1+xI(x))^2},$$
$$\frac{\partial}{\partial I'} S = (e^{-x} + \gamma e^{-Nx}) \frac{-x}{1+xI(x)},$$
$$\frac{\mathrm{d}}{\mathrm{d}x} \frac{\partial}{\partial I'} S = \frac{x(e^{-x} + \gamma N e^{-Nx})}{1+xI(x)} + (e^{-x} + \gamma e^{-Nx}) \frac{x^2 I'(x) - 1}{(1+xI(x))^2}.$$

Therefore, (54) simplifies to a linear equation which leads to the optimum interference function given in (46). ■

Figure (3) shows the capacity region for $N = 5$ and $\mathscr{P} = 100$. It is observed that the maximum $R_{ave}$ is achieved for $R_{mul} \leq 1.05$. It is shown in [21] that a good fraction of the highest expected rate with infinite layers is achieved by two layers. Figure (3) shows that this is true for our multicast network as well. Furthermore, we observe that the two-layer rate region gets closer to the capacity region for higher values of $R_{mul}$.

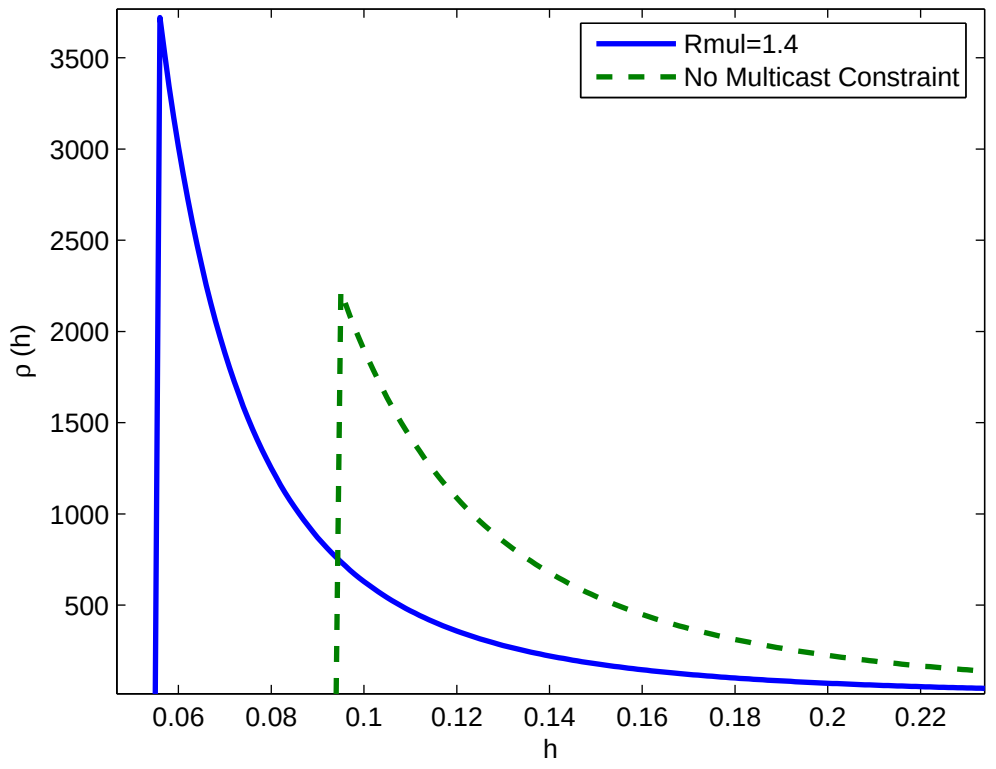

Fig. 4. Power distribution function $\rho(h)$ for no multicast requirement and for $R_{mul} = 1.4$

Without the multicast constraint, the power distribution function to maximize the average rate would be [5],

$$\rho_0(h) = \begin{cases} \frac{2}{h^3} - \frac{1}{h^2} & s_0 < h < 1 \\ 0 & else \end{cases},$$

where

$$s_0 = \frac{2}{1 + \sqrt{1 + 4\mathscr{P}}}.$$

This function is depicted in figure (4), and is compared with the case corresponding to a multicast requirement of $R_{mul} = 1.4$. As shown in the figure, the coverage requirement has shifted the power distribution to smaller channel gains to provide service to users with worse channel conditions.

## VI. Extension to MISO

In the case that there are multiple ($M > 1$) antennas at the transmitter, one can adopt the broadcast approach proposed in [5]. In this approach, the receiver with unknown quasi-static fading MISO channel, denoted by $\mathbf{h}$, is modeled as a continuum of receivers each associated with a channel realization. These receivers are ordered in a degraded fashion. However, since MIMO Broadcast Channel (MIMO-BC) is inherently non-degraded [22], this approach dose not necessarily lead to the optimum performance.

Assuming a single antenna at each receiver, the ordering of the virtual receivers in this approach is based on their normalized channel norm, i.e., $\frac{\|\mathbf{h}\mathbf{h}^{\dagger}\|}{M}$. Hence, the rate of the virtual receiver at the fading level $\frac{\|\mathbf{h}\mathbf{h}^{\dagger}\|}{M}$ equals

$$R\left(\frac{\|\mathbf{h}\mathbf{h}^{\dagger}\|}{M}\right) = \log\left(1 + \frac{P_S \frac{\|\mathbf{h}\mathbf{h}^{\dagger}\|}{M}}{1+P_I \frac{\|\mathbf{h}\mathbf{h}^{\dagger}\|}{M}}\right),$$

where $P_S$ and $P_I$ are the decodable and undecodable signal power levels, respectively.

Now, assume there are $N$ users in the network, all receiving a common source through an infinite-layer code. Same as before, we would like to design the code to maximize the average rate observed by a typical user, while providing a given coverage constraint for all the users. For this purpose, we should provide the minimum rate to the worst user in the degraded broadcast model, i.e., the user with the lowest channel norm. The normalized channel norm of user $i$, denoted by $\frac{\|\mathbf{h}_i\mathbf{h}_i^\dagger\|}{M}$, is a scaled $\chi^2$ random variable with $2M$ degrees of freedom, with the following CDF:

$$F_{typ}(h) = F_{\frac{\|\mathbf{h}\mathbf{h}^\dagger\|}{M}}(h) = 1 - \frac{\Gamma(M, Mh)}{\Gamma(M)}, \tag{55}$$

where $\Gamma(.)$ is the Gamma function, and $\Gamma(.,.)$ is the upper incomplete Gamma function [23]. Since the users' channels are statistically independent, the distribution of the norm of the worst channel can be computed as

$$\begin{aligned}\Pr\left\{\min_i \frac{\|\mathbf{h}_i\mathbf{h}_i^\dagger\|}{M} > h\right\} &= \left(\Pr\left\{\frac{\|\mathbf{h}_i\mathbf{h}_i^\dagger\|}{M} > h\right\}\right)^N \\ &= \left(\frac{\Gamma(M, Mh)}{\Gamma(M)}\right)^N.\end{aligned}$$

Hence, the CDF for the worst channel norm is

$$F_{mul}(h) = 1 - \left(\frac{\Gamma(M, Mh)}{\Gamma(M)}\right)^N. \tag{56}$$

Here, we just consider the soft coverage constraint scenario. Following the same approach as in section IV, we obtain

$$\begin{aligned}R_{mul} &= \int_0^\infty (1 - F_{mul}(u))\mathrm{d}R(u) \\ &= \int_0^\infty \left[\frac{\Gamma(M, Mu)}{\Gamma(M)}\right]^N \frac{u\rho(u)\mathrm{d}u}{1 + uI(u)},\end{aligned} \tag{57}$$

$$\begin{aligned}R_{ave} &= \int_0^\infty (1 - F_{typ}(u))\mathrm{d}R(u) \\ &= \int_0^\infty \frac{\Gamma(M, Mu)}{\Gamma(M)} \frac{u\rho(u)\mathrm{d}u}{1 + uI(u)},\end{aligned} \tag{58}$$

where $\rho(u)$ and $I(u)$ are the corresponding power allocation and interference power functions, respectively. To characterize the achievable rate region $(R_{ave}, R_{mul})$, we need to solve the following optimization problem for all $\gamma \geq 0$:

$$\max_{I(.)} R_{ave} + \gamma R_{mul}, \tag{59}$$

subject to

$$\begin{aligned}I(0) &= \mathscr{P}, \\ I(\infty) &= 0.\end{aligned}$$

Defining $S(x, I(x), I'(x), \gamma)$ as

$$\begin{aligned} S(x, I(x), I'(x), \gamma) &= \frac{\Gamma(M, Mx)}{\Gamma(M)} \frac{xI'(x)}{1 + xI(x)} \\ &+ \gamma \left( \frac{\Gamma(M, Mx)}{\Gamma(M)} \right)^N \frac{xI'(x)}{1 + xI(x)}, \end{aligned}$$

and setting its functional variation equal to zero to maximize the average rate, and defining

$$w(x) \triangleq \frac{\Gamma(M, Mx)}{\Gamma(M)} + \gamma \left( \frac{\Gamma(M, Mx)}{\Gamma(M)} \right)^N, \tag{60}$$

similar to (54), we obtain

$$\frac{x}{1 + xI(x)} w'(x) + \frac{w(x)}{(1 + xI(x))^2} = 0, \tag{61}$$

which implies that

$$I(x) = -\frac{w(x)}{x^2 w'(x)} - \frac{1}{x}. \tag{62}$$

Noting that $\frac{\mathrm{d}}{\mathrm{d}x}\Gamma(M, x) = -x^{M-1} e^{-x}$, we can write

$$w'(x) = -\frac{M^M x^{M-1} e^{-Mx}}{\Gamma(M)} \left( 1 + \gamma N \left[ \frac{\Gamma(M, Mx)}{\Gamma(M)} \right]^{N-1} \right). \tag{63}$$

Substituting in (62) yields the optimizer $I(h)$ as

$$I(h) = \begin{cases} \mathscr{P} & \text{if } h < h_0 \\ \mu(h) & h_0 < h < h_1 \\ 0 & h > h_1 \end{cases}, \tag{64}$$

where

$$\mu(h) \triangleq \frac{\Gamma(M, Mh) + \gamma \frac{\Gamma(M, Mh)^N}{\Gamma(M)^{N-1}}}{M^M h^{M+1} e^{-Mh} \left( 1 + \gamma N \left( \frac{\Gamma(M, Mh)}{\Gamma(M)} \right)^{N-1} \right)} - \frac{1}{h},$$

and $h_0$, $h_1$ are the solutions of the following equations:

$$\begin{aligned} \mu(h_0) &= \mathscr{P}, \\ \mu(h_1) &= 0, \end{aligned}$$

respectively.

The achievable rate region is shown in figure (5) for different values of $N$ and for $M = 2$. From this figure, we can see that as the number of users decreases, the proposed achievable rate region expands. It is also evident by comparing the regions for the MISO and SISO cases with $N = 5$ users (figures (3) and (5)) that using multiple antennas improves the achievable rate. However, its effect is more considerable for the multicast channel as compared to the typical channel. This prominent gain for the multicast channel is due to the fact that we are using multiple independent paths to convey data (higher diversity order), so the probability of having very low channel gains significantly decreases. In fact, we will show that one can compensate the effect of having a large number of users by increasing the number of transmit antennas. More specifically, if both $N$ and $M$ tend to infinity and $M$ grows fast enough with respect to $N$, we show in the next theorem that the multicast rate can reach the expected typical rate and our scheme gives the optimal solution.

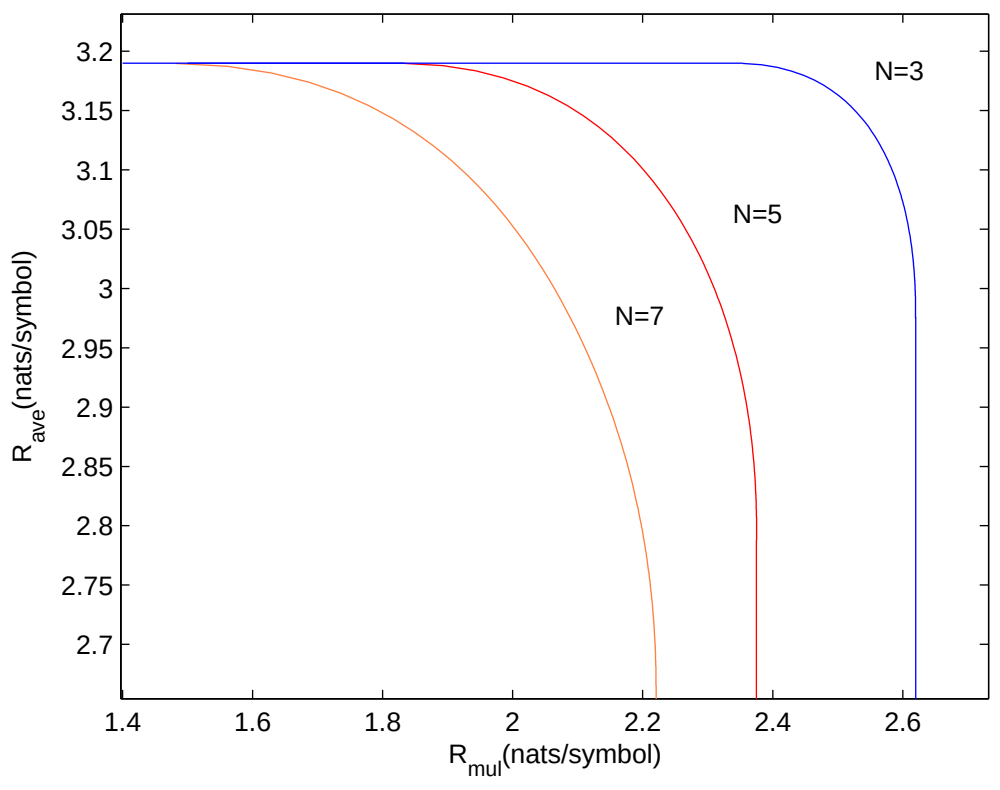

Fig. 5. Soft coverage constraint: MISO expected minimum rate vs. expected typical rate for different number of users, $M = 2$ and $\mathscr{P} = 100$

**Theorem 4** *For large values of $M$ and $N$, the proposed infinite layer superposition coding will provide $R_{mul}$ such that*

$$R_{mul} \geq R_{opt} - \sigma, \tag{65}$$

*if*

$$M > \frac{2\mathscr{P}^2 \log(N) + \omega(1)}{(1+\mathscr{P})^2\sigma^2}, \tag{66}$$

*where $R_{opt}$ is the highest achievable average rate for a typical user in the network, $\sigma$ is an arbitrarily small positive number, and $\omega(1)$ denotes any function of $N$ which tends to infinity as $N \to \infty$.*

*Proof:* First, we derive an upper bound on the achievable average rate for a typical user by assuming no stringent delay constraint, meaning that the transmission block can be chosen as large as the fading block. In this case, the channel has an ergodic behavior, and hence, the ergodic capacity is defined and is shown to be

$$C_{erg} = E\left[\log\left(1 + \frac{\|\mathbf{h}\mathbf{h}^{\dagger}\|}{M}\mathscr{P}\right)\right]. \tag{67}$$

As a result,

$$R_{opt} \leq C_{erg}. \tag{68}$$

Using the concavity of $\log$ function, and having the fact that $E\left[\frac{\|\mathbf{h}\mathbf{h}^{\dagger}\|}{M}\right]=1$, we have

$$C_{erg} \leq \log(1+\mathscr{P}). \tag{69}$$

We will show that our scheme provides a multicast rate arbitrarily close to this upper bound, if we use enough transmit antennas. Since this upper bound is larger than the expected typical rate, the theorem will be proved. Using a single-layer code with power $\mathscr{P}$ [7] and rate $R_{\sigma}$, where

$$R_{\sigma} = \log(1+\mathscr{P}(1-\sigma')), \tag{70}$$

and

$$\sigma' = \frac{(1+\mathscr{P})\sigma}{\mathscr{P}},$$

the expected multicast rate in our network will be

$$R_{mul} = \Pr\left\{\frac{\|\mathbf{h}\mathbf{h}^{\dagger}\|_{mul}}{M} > 1-\sigma'\right\} R_{\sigma}, \tag{71}$$

where $\|\mathbf{h}\mathbf{h}^{\dagger}\|_{mul} = \min_i \|\mathbf{h}_i\mathbf{h}_i^{\dagger}\|$. Regarding the central limit theorem [23], the distribution of $\frac{\|\mathbf{h}\mathbf{h}^{\dagger}\|}{M}$, where

$$\frac{1}{M}\|\mathbf{h}\mathbf{h}^{\dagger}\| = \frac{{h_1}^2 + {h_2}^2 + \ldots + {h_M}^2}{M}, \tag{72}$$

and $\{h_m\}_{m=1}^{M}$'s are independent Rayleigh distributions with unit variance and unit mean, approaches to a Gaussian distribution with the CDF

$$F(h) = Q\left(\frac{h-1}{\frac{1}{\sqrt{M}}}\right). \tag{73}$$

Consequently, the CDF of the multicast channel will be

$$F_{mul}(h) \approx 1 - Q\left(\frac{h-1}{\frac{1}{\sqrt{M}}}\right)^{N}. \tag{74}$$

Using the above equation, (71) can be written as

$$\begin{aligned} R_{mul} &\approx Q(-\sqrt{M}\sigma')^{N} R_{\sigma} \\ &= \left[1 - Q(\sqrt{M}\sigma')\right]^{N} R_{\sigma}. \end{aligned} \tag{75}$$

Assuming $M$ is large enough to have $\sqrt{M}\sigma' >> 1$, and consequently $Q(\sqrt{M}\sigma') << 1$, we can rewrite the above equation as

$$R_{mul} \approx e^{-NQ(\sqrt{M}\sigma')} R_{\sigma}. \tag{76}$$

[7]Note that as the single-layer coding is a special case of superposition coding, our proposed scheme outperforms the single-layer coding scheme.

Now, using the approximation

$$Q(x) \approx \frac{1}{\sqrt{2\pi x}} e^{-\frac{x^2}{2}} \tag{77}$$

for large values of $x$, we can write

$$Q(\sqrt{M}\sigma') \leq e^{-\frac{M\sigma'^2}{2}}. \tag{78}$$

Therefore, having

$$M = \frac{2\log(N) + \omega(1)}{\sigma'^2}, \tag{79}$$

incurs $NQ(\sqrt{M}\sigma') = o(1)$, and as a result,

$$\lim_{N\to\infty} R_{mul} - R_\sigma = 0. \tag{80}$$

Moreover, assuming $\sigma \ll 1$, (70) can be written as,

$$\begin{aligned} R_\sigma &\approx \log(1+\mathscr{P}) - \frac{\mathscr{P}\sigma'}{1+\mathscr{P}} \\ &\geq C_{erg} - \sigma, \end{aligned} \tag{81}$$

where the second line results from (69). Combining (68), (80), and (81), the result of Theorem 4 easily follows. ■

Theorem 4 simply implies that as the number of transmit antennas grows at least logarithmically with the number of users, the gain of the worst channel converges to the gain of the typical channel in the network, with probability one. In other words, increasing the number of transmit antennas provides fairness in the system, such that all users almost get the same quality of service. This fact is also noticed by [24] in the context of MIMO-BC.

## VII. Conclusion

In this paper, we have considered a multicast channel, where a common data is transmitted from a source to several users. It is assumed that a minimum service must be provided for all the users. For this setup, we have optimized the average service received by a typical user in the network. Two scenarios are considered for the coverage constraint. In the case of hard coverage constraint, the minimum multicast requirement is stated in terms of the minimum rate (multicast outage capacity) received by all the users in a single transmission block. For small enough outage probabilities, it is shown that the capacity region is achieved by providing the required multicast rate in a single layer code, and designing an infinite-layer code as in [5], on top of it. In the case of soft coverage constraint, the multicast requirement is expressed in terms of the expected minimum rate received by all the users. An infinite layer superposition coding is shown to achieve the capacity region in this scenario. We have also proposed a suboptimal coding scheme for the MISO multicast channel. This scheme is shown to be asymptotically optimal, when the number of transmit antennas grows at least logarithmically with the number of users.